%% file: main.tex
\documentclass[sigconf]{acmart}
\usepackage{booktabs}  
\usepackage{multirow}  
\usepackage{colortbl}  
\usepackage{xcolor}    
\usepackage{tabularx}
\usepackage{amsmath}
\usepackage{pgfplots}
\usepackage{tikz}
\pgfplotsset{compat=1.18}
\usepackage{graphicx}

\usepackage{amssymb}   
\usepackage{enumitem}
\usepackage{pifont}   
\usepackage{xcolor}
\usepackage[table,xcdraw]{xcolor}
\AtBeginDocument{%
  }

\copyrightyear{2026}
\acmYear{2026}
\acmConference[SIGIR '26]{Proceedings of the 49th International ACM SIGIR Conference on Research and Development in Information Retrieval}{July 20--24, 2026}{Melbourne, VIC, Australia}
\acmBooktitle{Proceedings of the 49th International ACM SIGIR Conference on Research and Development in Information Retrieval (SIGIR '26), July 20--24, 2026, Melbourne, VIC, Australia}

\begin{document}

\title{FollowTable: A Benchmark for Instruction-Following Table Retrieval}

\author{Rihui Jin}
\additionalaffiliation{%
\institution{Key Laboratory of New Generation Artificial Intelligence Technology and Its Interdisciplinary Applications (Southeast University), Ministry of Education}
\city{Nanjing}
\country{China}}
\affiliation{
  \institution{Southeast University}
  \city{Nanjing}
  \country{China}}
\email{230248998@seu.edu.cn}

\author{Yuchen Lu}
\authornotemark[1]
\affiliation{
  \institution{Southeast University}
  \city{Nanjing}
  \country{China}}
\email{220245051@seu.edu.cn}

\author{Ting Zhang}
\authornotemark[1]
\affiliation{
  \institution{Southeast University}
  \city{Nanjing}
  \country{China}}
\email{ting.zhang@seu.edu.cn}

\author{Jun Wang}
\authornotemark[1]
\affiliation{
  \institution{Southeast University}
  \city{Nanjing}
  \country{China}}
\email{220252392@seu.edu.cn}

\author{Kuicai Dong}
\affiliation{
  \institution{Huawei Noah’s Ark Lab}
  \city{Shenzhen}
  \country{China}}
\email{kuicai001@e.ntu.edu.sg}

\author{Zhaocheng Du}
\authornote{Corresponding author.}
\affiliation{
  \institution{Huawei Noah’s Ark Lab}
  \city{Shenzhen}
  \country{China}}
\email{zhaochengdu@huawei.com}

\author{Dongping Liu}
\authornotemark[1]
\affiliation{
  \institution{Southeast University}
  \city{Nanjing}
  \country{China}}
\email{213232400@seu.edu.cn}

\author{Gang Wang}
\affiliation{
  \institution{Huawei Noah’s Ark Lab}
  \city{Shenzhen}
  \country{China}}
\email{wanggang110@huawei.com}

\author{Yong Liu}
\affiliation{
  \institution{Huawei Noah’s Ark Lab}
  \city{Shenzhen}
  \country{China}}
\email{liu.yong6@huawei.com}

\author{Guilin Qi}
\authornotemark[1]
\affiliation{
  \institution{Southeast University}
  \city{Nanjing}
  \country{China}}
\email{gqi@seu.edu.cn}
\renewcommand{\shortauthors}{Rihui Jin et al.}

\begin{abstract}
  \input{sections/0.abstract}

\end{abstract}



\maketitle

\input{sections/1_intro}

\input{sections/2_related}

\input{sections/3_formulation}

\input{sections/4_benchmark}

\input{sections/6_experiment1}
\input{sections/6_experiment2}
\input{sections/8_conclusion}

\begin{acks}
This work is partially supported by National Nature Science Foundation of China under No. 62476058. We thank the Big Data Computing Center of Southeast University for providing the facility support on the numerical calculations in this paper.
\end{acks}

\bibliographystyle{ACM-Reference-Format}
\balance
\bibliography{citation}

\end{document}

%% file: sections/0.abstract.tex
Table Retrieval (TR) has traditionally been formulated as an ad-hoc retrieval problem, where relevance is primarily determined by topical semantic similarity.
With the growing adoption of LLMs-based agentic systems, access to structured data is increasingly instruction-driven, where relevance is conditional on explicit content and schema constraints rather than topical similarity alone.
We therefore formalize Instruction-Following Table Retrieval (IFTR), a new task that requires models to jointly satisfy topical relevance and fine-grained instruction constraints.
We identify two core challenges in IFTR: (i) sensitivity to content scope, such as inclusion and exclusion constraints, and (ii) awareness of schema-grounded requirements, including column semantics and representation granularity—capabilities largely absent in existing retrievers.
To support systematic evaluation, we introduce \textbf{\textsc{FollowTable}}, the first large-scale benchmark for IFTR, constructed via a taxonomy-driven annotation pipeline.
We further propose a new metric, termed the Instruction Responsiveness Score , to evaluate whether retrieval rankings consistently adapt to user instructions relative to a topic-only baseline.
Our results indicate that existing retrieval models struggle to follow fine-grained instructions over tabular data.
In particular, they exhibit systematic biases toward surface-level semantic cues and remain limited in handling schema-grounded constraints, highlighting substantial room for future improvements.
Our benchmark is released publicly.\footnote{\url{https://github.com/AriKing11/FollowTable_Benchmark}}

%% file: sections/1_intro.tex
\section{Introduction}
\label{sec:introduction}
Table Retrieval (TR)~\cite{sig_table_search_zhiyu, DBLP:conf/aidm/BodensohnB24, gnqtables} is a fundamental task in Information Retrieval (IR) that identifies relevant tables from large-scale repositories to satisfy user needs. 
Traditionally, TR has been formulated as a conventional ad-hoc retrieval problem~\cite{wtr}, primarily focused on semantic matching to support downstream applications such as Open Table Question Answering~\cite{nq_tables} or Data Integration~\cite{zhu2025large, zhang_survey}. 
However, the proliferation of LLMs and intelligent agents~\cite{IR_Llm_agent, DBLP:conf/sigir/TianTWYP25,DBLP:conf/sigir/ArabzadehCPSST25} has fundamentally redefined user expectations, shifting the paradigm from simple query-based search toward complex, instruction-driven access to structured knowledge, where tables serve as a rich information source in sophisticated applications such as Multi-modal Retrieval Augmented Generation~\cite{xu2025mmrag,mrag_survey,ott_qa}.

\begin{figure}
    \centering
    \includegraphics[width=0.88\linewidth]{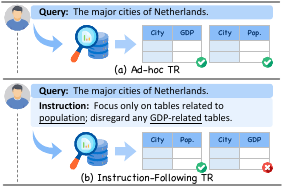}
    \vspace{-0.2cm}
    \caption{Ad-hoc (a) v.s. instruction-following (b) table retrieval. In (a), topic similarity suffices for relevance (\ding{51}). In (b), tables are rejected (\ding{55}) if they do not satisfy the specific constraints defined in the instruction.}
    \vspace{-0.5cm}
    \label{fig:intro}
\end{figure}

In this evolving landscape, relevance in tabular retrieval with user inputs is no longer an intrinsic property determined solely by topical overlap (Fig.~\ref{fig:intro}(a)), but is instead conditioned on explicit, fine-grained instructions that constrain table content scope, schema attributes, and structural granularity (Fig.~\ref{fig:intro}(b)).
As a consequence, tables that are topically aligned with a query may still be irrelevant if they violate specific logical or structural constraints, exposing a critical gap in current IR systems.
We argue that this gap stems from a fundamental limitation of existing dense retrievers, which are predominantly optimized for holistic semantic similarity and thus suffer from a severe representation bottleneck when handling such multi-faceted constraints.

Specifically, this limitation manifests in two fundamental challenges unique to the intersection of instructions and tabular data:
(1) \textbf{Sensitivity to Table Content Scope Constraints:} Unlike ad-hoc TR, instructions often define strict search boundaries involving inclusion, exclusion, or exclusive filtering.
Current retrievers, mainly driven by global semantic overlap, exhibit a semantic bias where they struggle to distinguish between a mentioned entity and a logically required or forbidden one. 
(2) \textbf{Awareness of Table Schema Constraints:} Instructions frequently impose requirements on table-specific schema, such as specific column semantics or data granularity. 
Existing models, which naively linearize or flatten tables into unstructured sequences, lack explicit mechanisms to align these schema-level instructions with the underlying tabular data.

\begin{figure*}
    \centering
    \includegraphics[width=0.95\linewidth]{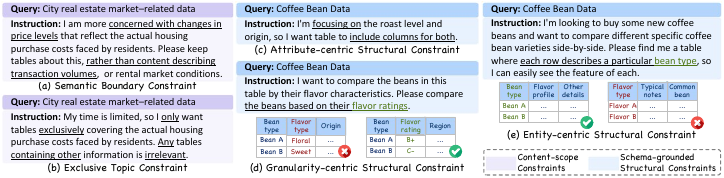}
    \vspace{-0.2cm}
    \caption{Illustration of the proposed taxonomy for IFTR.
Instructions are divided into Content-scope Constraints (purple) and Schema-grounded Structural Constraints (blue), covering five fine-grained types.
In the harder-to-interpret cases (d) and (e), example tables are provided to show that, despite being query-relevant, only the table marked with \ding{51} follows the instruction, whereas \ding{55} indicates an instruction violation.}
    \label{fig:intro_eg}
    \vspace{-0.2cm}
\end{figure*}

Motivated by these observations, we propose that TR must evolve from semantic matching-only toward \textbf{Instruction-Following Table Retrieval (IFTR)}. 
IFTR requires models to jointly satisfy fine-grained user instructions and topical relevance, leading to instruction-conditioned relevance judgments that fundamentally go beyond ad-hoc retrieval with longer queries.
To bridge this gap, we introduce \textbf{\textsc{FollowTable}}, the first large-scale benchmark for IFTR. 
Constructed via a rigorous annotation pipeline and grounded in our proposed taxonomy, it enables systematic and fine-grained evaluation of instruction-following capabilities across diverse constraint types.
Our taxonomy categorizes instructions into: 
(i) \textit{Content-scope Constraints}, covering boundary and exclusion requirements; and 
(ii) \textit{Schema-grounded Structural Constraints}, addressing attribute column focus, entity column centering, and column representation granularity.
Each of these two high-level categories is further decomposed into five fine-grained constraint types, with representative examples illustrated in Fig.~\ref{fig:intro_eg}.
To address the limitations of existing IR metrics, we propose the Instruction Responsiveness Score (IRS), 
a set-level metric that quantifies instruction-induced ranking changes relative to a topic-only baseline.

Extensive experiments on \textsc{FollowTable} reveal the significant challenges inherent in IFTR for current retrievers. 
In addition, while general re-rankers outperform the best dense retriever by approximately 16 points in IRS, this improvement comes at the cost of nearly 100$\times$ higher inference latency. 
Crucially, our analysis uncovers a pervasive ``Positive Attention Bias,'' where models paradoxically promote tables containing explicitly negated entities, ignoring exclusionary constraints. 
Moreover, we observe that schema-grounded structural constraints pose a significantly greater challenge to current models than content-scope constraints. 
Finally, empirical validation confirms that our proposed IRS metric provides a more monotonic and sensitive signal of instruction adherence than traditional metrics like nDCG, which often fail to capture the conditional nature of relevance in IFTR.

In summary, our contributions are as follows:
\begin{itemize}
    \item We formulate IFTR, a new IR task that requires conditional relevance judgments under user instructions, extending conventional ad-hoc TR.
    \item We introduce \textsc{FollowTable}, the first large-scale benchmark for IFTR, featuring instruction-annotated tables grounded in a taxonomy of table-specific constraints, and accompanied by a novel evaluation metric IRS.
    \item We provide extensive experimental analyses revealing fundamental limitations of current retrievers in handling instructions containing content-scope and schema-grounded constraints, offering insights for future TR research.
\end{itemize}

%% file: sections/2_related.tex
\section{Related Work}

\subsection{Table Retrieval}
\label{sec:related_tr}
TR has become a pivotal research area within IR, driven by the need to handle complex structured data~\cite{miutra2018introduction, zhang_survey}. 
While large-scale web table datasets~\cite{WebQueryTable, WikiTables_60, table_Retrieval_Solved_Problem} and domain-specific benchmarks~\cite{TableArXiv, ji2025target} have established standard evaluation protocols, TR has gained renewed importance in the context of Hybrid RAG systems. 
In these settings, benchmarks like~\cite{nqtables, ott_qa, xu2025mmrag} emphasize the critical role of TR in integrating structured tabular knowledge with generative models to improve factual accuracy.

Methodologically, TR approaches~\cite{Table_Pre_training} are divided into two categories. 
The first involves structure-aware models that explicitly encode cell-level semantics and row/column dependencies~\cite{gtr, nq_tables, jin2023tabprompt, strbert, SSDR, ecat, thyme}. 
For instance, Tapas-DTR~\cite{nq_tables} jointly processes table structures and query context for fine-grained matching. 
The second category simplifies tables through linearization with some additional schema tokens, applying generic text retrieval frameworks without any design specifically encoding table structure~\cite{min2025unihgkr, min2024exploring, table_Retrieval_Solved_Problem, TR_Necessitate, utp, Birdie}.
Notably, these approaches have demonstrated significant efficacy, leveraging the powerful representation and matching capabilities of modern general-purpose retrievers to exhibit superior zero-shot performance on tables.

Despite this progress, current research fails to explicitly evaluate or model the ability of TR systems to adhere to the instructions, creating a mismatch between benchmark performance and the demands of real-world deployment.

\subsection{Instruction-Following Information Retrieval}
\label{sec:related_ifir}
The evolution of IR has witnessed a paradigm shift from semantic similarity matching toward fine-grained modeling of nuanced user intents. 
Early benchmarks~\cite{petroni2021kilt, muennighoff2023mteb}, such as  BEIR~\cite{thakur2021beir} established the foundation for IFIR evaluation. 
However, these benchmarks typically rely on static, dataset-level instructions, which fail to capture the query-specific constraints prevalent in real-world scenarios. 
To address this, recent benchmarks~\cite{InfoSearch, sun2024mair, weller2025followir, oh2024instructir} have introduced instructions coupled with queries to evaluate rigorous adherence. 
For example, FollowIR~\cite{weller2025followir} pioneered this direction by repurposing TREC narratives into explicit constraints.

Parallel to these evaluation shifts, IR methods have evolved from BERT-based discriminative encoders to LLM-based generative architectures.
Early approaches like INSTRUCTOR~\cite{one_embedder} incorporated task instructions into BERT-style encoders but struggled with complex constraints due to limited model capacity.
To overcome this, recent research has pivoted to the LLM-based paradigm, leveraging decoder-only architectures for dense retrieval~\cite{zuccon2025r2llms_tutorial}.
Models such as GritLM~\cite{gritlm}, RepLLaMA~\cite{repllama}, and Promptriever~\cite{Promptriever} repurpose generative decoders to derive semantically rich embeddings, exhibiting superior instruction-following capabilities.

Despite these advancements, current IFIR research remains focused on unstructured text, overlooking the unique structural challenges of tabular data as mentioned in \S\ref{sec:introduction}.

%% file: sections/3_formulation.tex
\section{Problem Formulation}
\label{sec:formulation}

This section formally introduce the notations used in this paper.

\subsection{Ad-hoc TR}
We consider TR tasks where a user issues a NL query to retrieve relevant tables from a corpus~\cite{wtr}. Let $\mathcal{C}_t=\{t_1,t_2,\dots,t_M\}$ denote a corpus of $M$ tables, where each table $t$ consists of structured tabular content along with optional metadata. Let $\mathcal{Q}$ denote the set of all queries. 
In the conventional setting, each data instance is defined as $d = (q, \mathcal{T}^{+}_{q})$, where $q \in \mathcal{Q}$ represents a user query expressing a coarse-grained information need, and $\mathcal{T}^{+}_{q} \subseteq \mathcal{C}_t$ denotes the set of tables relevant to $q$. Tables not annotated as relevant are defined as $\mathcal{T}^{-}_{q}=\mathcal{C}_t\setminus\mathcal{T}^{+}_{q}$, which represents the set of irrelevant (negative) tables. 
Retrieval models in this context aim to estimate query--table relevance, typically via semantic matching, without additional constraints. 
Given a query $q \in \mathcal{Q}$, a table retrieval model produces a ranked list
$\pi_q = (t_{(1)}, t_{(2)}, \dots, t_{(K)})$,
where $t_{(i)} \in \mathcal{C}_t$ denotes the table ranked at position $i$, and $K \le M$ is the retrieval cutoff.
The ranked list $\pi_q$ is typically induced by a scoring function
$f(q, t): \mathcal{Q} \times \mathcal{C}_t \rightarrow \mathbb{R}$,
such that tables are ordered in descending order of their relevance scores.
An ideal ranked list should satisfy:
\[
\forall\, t^{+} \in \mathcal{T}^{+}_{q},\; t^{-} \in \mathcal{T}^{-}_{q}:
\quad f(q, t^{+}) > f(q, t^{-}),
\]
so that the ranked list $\pi_q$ contains as many positive tables as possible in top-ranked positions and minimizes the presence of negative tables among the early ranks.

\subsection{IFTR}
IFTR extends the traditional formulation by explicitly conditioning retrieval on an \emph{instruction} $i \in \mathcal{I}$ that refines the user intent. 
For each query $q \in \mathcal{Q}$, there exists a set of instructions $\mathcal{I}(q) = \{i_1, \dots, i_L\}$, where $\mathcal{I}(q) \subseteq \mathcal{I}$. A data instance is defined as $d=(q, i, \mathcal{T}^{+}_{qi}, \mathcal{T}^{-}_{qi})$, where $i \in \mathcal{I}(q)$. 
Here, $\mathcal{T}^{+}_{qi} \subseteq \mathcal{C}_t$ is the set of instruction-compliant positive tables, which are both relevant to $q$ and satisfy the constraints in $i$. 
In contrast, $\mathcal{T}^{-}_{qi} = \mathcal{T}^{+}_{q} \setminus \mathcal{T}^{+}_{qi}$ denotes the instruction-violating tables, which remain relevant to $q$ but fail to comply with $i$. 
Given a query--instruction pair $(q,i)$, an IFTR model produces an instruction-aware ranked list
$\pi_{q,i} = (t_{(1)}, t_{(2)}, \dots, t_{(K)})$,
where $t_{(i)} \in \mathcal{C}_t$ and $K \le M$.
The ranking is induced by an instruction-conditioned scoring function
$f(q,i,t): \mathcal{Q} \times \mathcal{I} \times \mathcal{C}_t \rightarrow \mathbb{R}$.
The retrieval objective is to rank instruction-compliant tables in $\mathcal{T}^{+}_{qi}$ ahead of instruction-violating tables in $\mathcal{T}^{-}_{qi}$ or $\mathcal{T}^{-}_{q}$, i.e.,
\[
\forall\, t^{+} \in \mathcal{T}^{+}_{qi},\; t^{-} \in \mathcal{T}^{-}_{qi} \cup \mathcal{T}^{-}_{q}:
\quad f(q,i,t^{+}) > f(q,i,t^{-}).
\]

%% file: sections/4_benchmark.tex
\section{\textsc{FollowTable} Benchmark}
\label{sec:benchmark}

\begin{figure}
    \centering
    \includegraphics[width=0.97\linewidth]{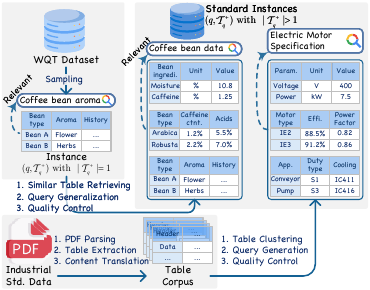}
    \vspace{-0.1cm}
    \caption{The data pre-processing and preparation pipeline for WQT expansion and \textsc{InduSTR} construction. It illustrates the workflow of transforming raw data into standardized $(q, \mathcal{T}^{+}_{q})$ with $|\mathcal{T}^{+}_{q}| > 1$ instances through generation, filtering, and quality control.}
    \label{fig:pre_pro} 
    \vspace{-0.3cm}
\end{figure}

\begin{table}[t]
    \input{tables/old_stats}
\end{table}

To bridge the gap between semantic relevance and instruction compliance, we introduce \textsc{FollowTable}, the first comprehensive benchmark designed specifically for IFTR. 
Unlike existing datasets that focus solely on topical matching, \textsc{FollowTable} requires models to satisfy explicit constraints across both content scopes and schema. 
In this section, we detail the construction process of this benchmark.
We begin by describing the data acquisition and pre-processing steps in \S\ref{sec:data_collection}. 
Next, we present our automated pipeline for generating diverse instructions in \S\ref{sec:instruction_gen}. 
We then define rigorous metrics to evaluate instruction adherence in \S\ref{sec:irs}, and finally, we provide a statistical analysis of the dataset in \S\ref{sec:dataset_analysis}. 
During the data generation and annotation process, it is all handled by LLMs\footnote{The LLMs in this chapter refers to GPT-5.2.}. In the quality review stage, both LLMs and humans are involved.

\subsection{Data Collection and Preparation}
\label{sec:data_collection}

\subsubsection{Data Collection}

To construct instruction-following instances for TR, we require a setup where each query is associated with multiple relevant tables ($|\mathcal{T}^{+}_{q}| > 1$). This structure allows instructions to meaningfully refine the target table set.
Existing open-domain table QA~\cite{nq_tables} or RAG datasets~\cite{min2025unihgkr, ott_qa} are unsuitable for this purpose due to:
1) the lack of explicitly annotated irrelevant tables (negatives), or
2) relevance annotations that only cover partial sub-facts of multi-hop queries rather than full relevance.

To address this, we adapt three established TR datasets (summarized in Table~\ref{tab:dataset_statistics}) and introduce a new source: Industrial Standard Manuals.
These manuals, collected as 5,000 PDF documents, contain hierarchically structured tables across diverse domains (e.g., chemical engineering, manufacturing).
Crucially, tables within the same chapter often describe closely related specifications, providing a rich source of semantically similar table clusters. We denote this newly constructed subset as \textsc{IndusTR}.

\subsubsection{Data Preparation}
\label{subsec:Data_Preparation}

Our objective is to unify all sources into the standard input format $(q, \mathcal{T}^{+}_{q})$, ensuring $|\mathcal{T}^{+}_{q}| > 1$.
While WTR and TableArXiv already align with this format, WQT and \textsc{InduSTR} require specific processing pipelines, as illustrated in Fig.~\ref{fig:pre_pro}.

\textbf{WQT Expansion.}
The original WQT pairs each query $q$ with a single positive table. To enable refinement, we expand the relevant set $\mathcal{T}^{+}_{q}$ via a generalize-and-filter approach:
\begin{enumerate}
    \item \textbf{Similar Table Retrieving:} We retrieve a set of candidate tables semantically similar to the original positive table $t^{+}_{q}$ using dense embeddings.
    \item \textbf{Query Generalization \& Filtering:} We prompt an LLM to generalize $q$ into a broader query $q'$. The LLM then identifies tables from the candidates that satisfy $q'$, forming the expanded set $\mathcal{T}^{+}_{q'}$ with $|\mathcal{T}^{+}_{q'}| > 1$.
    \item \textbf{Quality Control:} We employ an LLM-based Quality Critic to validate both the quality of the query generalization and the accuracy of the table relevance annotations, discarding instances that fail this dual verification.
\end{enumerate}

\textbf{\textsc{IndusTR} Construction.}
We construct instances from the raw industrial PDFs in a two-phase process:
Phase 1: Corpus Construction. We first build a standardized table corpus. We employ MinerU~\cite{wang2024mineruopensourcesolutionprecise} to parse raw PDF documents into HTML format and extract tables using heuristic rules. 
Subsequently, we utilize an LLM to translate all English tables into Chinese. To ensure technical precision, we extract domain-specific bilingual glossaries from the source documents and inject them as reference context during the translation process.
Phase 2: Instance Generation. We then transform this corpus into standard input instances using a pipeline analogous to WQT, with specific adaptations:
\begin{enumerate}
    \item \textbf{Table Clustering:} Cluster the tables based on semantic similarity and iteratively select one cluster as the candidates.
    \item \textbf{Query Generation \& Filtering:} For each selected cluster, the LLM generates a new query $q$ that describes the shared semantics of the tables (rather than generalizing an existing query). The LLM then filters the cluster to retain only tables strictly relevant to $q$, forming $\mathcal{T}^{+}_{q}$.
    \item \textbf{Quality Control:} We apply the same dual verification mechanism as in WQT to ensure the consistency between the generated query and the table set.
\end{enumerate}

This pipeline lays the foundation for constructing a benchmark in which each query is associated with multiple relevant tables, enabling subsequent instruction-based refinement of relevance.

\subsection{Instruction Generation Pipeline}
\label{sec:instruction_gen}
\begin{figure*}
    \centering
    \includegraphics[width=0.97\linewidth]{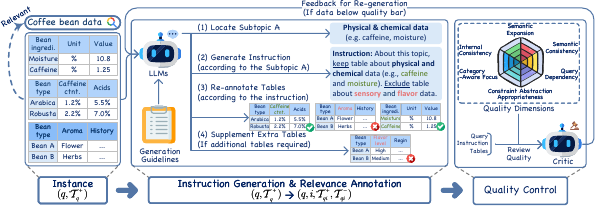}
    \vspace{-0.2cm}
    \caption{The automated instruction generation and quality review pipeline for \textsc{FollowTable}. The process transforms raw query-table pairs into instruction-following instances through four generation steps (subtopic mining, instruction generation, re-annotation, and supplementation) followed by a feedback-driven quality review using a Critic to evaluate six key dimensions.}
    \label{fig:benchmark}
    \vspace{-0.2cm}
\end{figure*}
To transform the collected query-table pairs into instruction following instances, we design an automated generation pipeline driven by LLMs.
Formally, given an instance $(q, \mathcal{T}^{+}_{q})$, our goal is to generate  natural language instruction $\mathcal{I}(q) = \{i_1, \dots, i_L\}$ and the refined ground-truth labels $\mathcal{T}^{+}_{qi}$ about instruction-table relevance.

The pipeline consists of four generation steps followed by a feedback-driven quality control stage:
\begin{enumerate}[leftmargin=*, label=\textbf{\arabic*.}, nosep]
\item \textbf{Feasibility Analysis \& Sub-topic Mining.}
The LLM first analyzes the semantic diversity within the table set $\mathcal{T}^{+}_{q}$.
It determines whether the tables can be logically divided based on specific concepts (e.g., temporal granularity, entity types, or specific metrics).
If the tables are too homogeneous to support a meaningful distinction via instructions, the instance is discarded.

\item \textbf{Category-Specific Instruction Generation.}
Based on the feasibility analysis, we prompt the LLM to generate an instruction $i$ that instantiates one of the fine-grained constraint categories defined in our taxonomy.
Specifically, we divide constraints into content-scope and schema-grounded structural categories, corresponding respectively to what semantic content the table should focus on and how that content is organized. It can be further categorized into the following types (Examples of each category are shown in Fig.~\ref{fig:intro_eg}):
    \begin{itemize}[leftmargin=1.5em, nosep]
        \item \textbf{Content-scope Constraints:}
        \begin{itemize}
            \item \textit{Semantic Boundary Constraint (C1):} The instruction narrows the topic to a clearly defined semantic subset. 
            \item \textit{Exclusive Topic Constraint (C2):} The instruction requires that tables concentrate exclusively on a particular aspect of the topic, without mixing in additional content.
        \end{itemize}
        
        \item \textbf{Schema-grounded Structural Constraints:}
        \begin{itemize}
            \item \textit{Attribute-centric Structural Constraint (S1):} The instruction requires the presence of attribute columns that describe specific properties or measurements of the entities, rather than merely matching a particular column name. 
            \item \textit{Entity-centric Structural Constraint (S2):} The instruction constrains which real-world entities the table is organized around. In such cases, the table is expected to be centered on a primary entity column, with the remaining columns representing attributes of that entity.
            \item \textit{Granularity-centric Structural Constraint (S3):} The instruction specifies the granularity of aggregation at which entity attributes are reported.
        \end{itemize}
    \end{itemize}

\item \textbf{Relevance Re-annotation.}
Given the generated instruction $i$, the relevance of each table in the query-positive set $\mathcal{T}_q$ is re-evaluated.
The LLM acts as an annotator, determining for each table $t \in \mathcal{T}_q$ whether it satisfies both the topical requirements of $q$ and the constraints of $i$.
This splits the tables into a new positive set $\mathcal{T}^{+}_{qi}$ and a negative set $\mathcal{T}^{-}_{qi}$.

\item \textbf{Instance Balancing \& Augmentation.}
To ensure the difficulty of the benchmark, we enforce a strict requirement: each valid instance must contain at least three positive tables ($|\mathcal{T}^{+}_{qi}| \ge 3$) and three hard negative tables ($|\mathcal{T}^{-}_{qi}| \ge 3$).
The negatives must be topically relevant to $q$ but violate $i$.
If an instance lacks sufficient positives or negatives after splitting, LLMs will synthesize more corresponding tables until the balance requirement is met.

\item \textbf{Feedback-Driven Quality Control via LLM-as-a-Judge.}
Generation alone does not guarantee instruction quality.
To ensure that synthetic instructions exhibit strict logical validity and realistic user intent, we introduce a multi-stage quality control mechanism in which evaluation results explicitly serve as feedback for regeneration.
Inspired by LLM-As-a-Judge~\cite{llm_as_a_judge}, we employ a separate LLM, acting as a \textit{Critic}, to evaluate each generated $(q, i)$ pair across the six rigorous dimensions summarized in Table~\ref{tab:evaluation_dimensions}.
Each dimension is rated on a three-level scale: \textit{Excellent}, \textit{Acceptable}, or \textit{Poor}.
If an instruction is rated as \textit{Poor} on any dimension, it is rejected and regenerated using the Critic’s feedback, forming a closed-loop refinement process.

\end{enumerate}

\begin{table}[]
    \input{tables/dimensions}
\end{table}

\begin{table*}[ht!]
    \input{tables/new_stats}
\end{table*}
\subsection{Evaluation Metrics for \textsc{FollowTable}}
\label{sec:irs}

While traditional metrics like nDCG~\cite{nDCG} assess topical relevance, recent work has introduced specialized metrics such as p-MRR~\cite{weller2025followir}, WISE~\cite{InfoSearch}, and InstFol~\cite{song2025ifir} to evaluate instruction compliance. 
However, these metrics are either operationally restrictive or fail to capture the nuances of TR.
Both WISE and InstFol are ill-suited for \textsc{FollowTable} due to their restrictive prerequisites or operational costs.
Specifically, WISE and InstFol are ill-suited for the \textsc{FollowTable} setting: WISE requires contrastive (paired) instructions which are not always available in general benchmarks, while InstFol's reliance on LLM-based judging incurs prohibitive costs and introduces stochastic variability that undermines reproducibility. 
In contrast, p-MRR serves as a proxy for instruction sensitivity but exhibits limited discriminative power; its scores often fluctuate irregularly and fail to distinguish between varying degrees of compliance. 
Most critically, p-MRR fails when the topic-only baseline is already optimal: in such scenarios, maintaining the perfect ranking yields a zero score for p-MRR because no rank shift is observed, thereby failing to acknowledge full instruction satisfaction.

To remedy these limitations, we propose the Instruction Responsiveness Score (IRS), a set-level metric specifically designed to quantify the degree to which an IFTR model reorders a list to satisfy instruction $i$ relative to a topic-only baseline.
For a query-instruction pair $(q,i)$, let $\pi_{q,i}$ be the instruction-aware ranking and $\pi_q$ be the original baseline ranking
induced by $f(q,t)$ (i.e., without conditioning on $i$).
Let $r(t,\pi)$ denote the 1-based rank of table $t$ in ranking $\pi$.
We use a rank-biased weight
\begin{equation}
w(r)=\frac{1}{\log_2(r+1)}.
\end{equation}
We then define the accumulated weighted mass of instruction-compliant and instruction-violating tables as
\begin{equation}
G^{+}(\pi)=\sum_{t\in \mathcal{T}^{+}_{qi}} w\!\left(r(t,\pi)\right),\quad
G^{-}(\pi)=\sum_{t\in \mathcal{T}^{-}_{qi}} w\!\left(r(t,\pi)\right).
\end{equation}
The unnormalized responsiveness score is
\begin{equation}
S(\pi_{q,i},\pi_q)=
\Bigl(G^{+}(\pi_{q,i})-G^{+}(\pi_q)\Bigr)-
\Bigl(G^{-}(\pi_{q,i})-G^{-}(\pi_q)\Bigr).
\end{equation}
This rewards promoting tables in $\mathcal{T}^{+}_{qi}$ and demoting tables in $\mathcal{T}^{-}_{qi}$
relative to $\pi_q$ (tables in $\mathcal{T}_q^{-}$ are treated as neutral and do not directly contribute).
To normalize across queries/instructions, we construct two reference rankings:
$\pi^{\mathrm{ideal}}_{q,i}$, which ranks tables in $\mathcal{T}^{+}_{qi}$ at the top
and places tables in $\mathcal{T}^{-}_{qi}$ at the bottom of the list; and $\pi^{\mathrm{worst}}_{q,i}$, which prioritizes $\mathcal{T}^{-}_{qi}$.
The final IRS is defined as
\begin{equation}
\mathrm{IRS}(\pi_{q,i})=
\begin{cases}
\dfrac{S(\pi_{q,i},\pi_q)}
{S(\pi^{\mathrm{ideal}}_{q,i},\pi_q)}, & S(\pi_{q,i},\pi_q)\ge 0,\\[12pt]
\dfrac{S(\pi_{q,i},\pi_q)}
{\bigl|S(\pi^{\mathrm{worst}}_{q,i},\pi_q)\bigr|}, & S(\pi_{q,i},\pi_q)< 0.
\end{cases}
\end{equation}
By construction, $\mathrm{IRS}\in[-1,1]$: $1$ indicates perfect instruction alignment,
$-1$ indicates maximal instruction violation, and $0$ means no change from the baseline.
When $\mathcal{T}^{+}_{qi}=\emptyset$ and $\mathcal{T}^{-}_{qi}=\emptyset$, we set $\mathrm{IRS}=0$.

\subsection{Dataset Analysis}
\label{sec:dataset_analysis}

\subsubsection{Quality Assurance}
We ensure the integrity of IFTR through a rigorous human-in-the-loop validation pipeline. 
First, every instruction is checked by two independent annotators against the six criteria in Table~\ref{tab:evaluation_dimensions}, with sub-standard ones revised or removed. 
To further validate the instruction-following relevance labels, we audit a subset of five tables per instruction, judging their alignment with both query topics and constraints. 
This comprehensive oversight yields a high inter-annotator agreement of \textbf{87\%} (Cohen’s $\kappa \approx \textbf{0.73}$), confirming the reliability of our data generation and annotation. 

\subsubsection{Data Characteristics}
Table~\ref{tab:new_stats} summarizes the key characteristics of \textsc{FollowTable} across four data sources.
Overall, \textsc{FollowTable} spans diverse domains and table structures, with substantial variation in table size and schema complexity, ensuring broad coverage beyond homogeneous benchmarks.
Each query is associated with a large candidate table set and multiple relevant tables, providing sufficient headroom for instruction-based refinement.
Across all subsets, instructions are well-balanced across content-scope and schema-grounded categories, with no single type dominating the benchmark, enabling fine-grained evaluation of different instruction-following behaviors.

%% file: tables/old_stats.tex
\centering
\caption{Statistics of table retrieval datasets adapted in this work.
 $|\mathcal{Q}|$ denotes the number of queries.
$|\mathcal{T}_\mathcal{Q}|$ refers to the total number of annotated tables including both relevant and irrelevant ones, as distinguished from the raw table corpus $C_t$.
$|\mathcal{T}_{\mathcal{Q}}^{+}|/|\mathcal{Q}|$ indicates the average number of relevant (positive) tables per query.}
\vspace{-0.2cm}
\resizebox{\linewidth}{!}{%
\begin{tabular}{lrrrrr}
\toprule
 Dataset                &  Main Domain  &  $|\mathcal{Q}|$          & $|\mathcal{T}_\mathcal{Q}|$      &  $|\mathcal{T}_{\mathcal{Q}}^{+}|/|\mathcal{Q}|$ \\
\midrule
WQT~\cite{WebQueryTable}        & Wikipedia             & 4{,}259              & 77k+                           & 1  \\
TableArXiv~\cite{TableArXiv}    & Physics               & 105                   & 8k+                            & 15 \\
WTR~\cite{wtr}                  & General               & 60                    & 6k+                             & 37 \\
\bottomrule
\end{tabular}
}
\vspace{-0.5cm}
\label{tab:dataset_statistics}

%% file: tables/dimensions.tex
\centering
\caption{Evaluation Dimensions for Instruction Quality.}
\vspace{-0.1cm}
\label{tab:evaluation_dimensions}
\small
\renewcommand{\arraystretch}{1.3} 

\newcolumntype{Z}{>{\raggedright\arraybackslash}m{0pt}<{\centering}} 
\renewcommand{\tabularxcolumn}[1]{m{#1}} 

\begin{tabularx}{\linewidth}{>{\raggedright\arraybackslash}m{1.95cm} X}
\toprule
\textbf{Dimension} & \textbf{Criterion \& Failure Modes} \\
\midrule

\textbf{Semantic \newline Expansion} & 
Introduction of non-trivial constraints to narrow retrieval space. \textit{Avoid: Decorative or vacuous constraints.} \\
\midrule

\textbf{Semantic \newline Consistency} & 
Logical refinement of the query without topic shifts. \textit{Avoid: Contradictions or unrelated topics.} \\
\midrule

\textbf{Query \newline Dependency} & 
Meaningful reliance on query context. \textit{Avoid: Verbatim repetition that renders the query redundant.} \\
\midrule

\textbf{Constraint Abstraction \newline Appropriateness} & 
Use of generalized conceptual/categorical constraints. \textit{Avoid: Overfitting or ``fingerprinting'' specific table values.} \\
\midrule

\textbf{Category-Aware \newline Focus} & 
Alignment between constraint type and semantic focus (content vs. schema). \textit{Avoid: Focusing on content for schema-type instructions, and vice versa.} \\
\midrule

\textbf{Internal \newline Consistency} & 
Absence of logical self-contradictions. \textit{Avoid: Simultaneously requiring and excluding the same property.} \\

\bottomrule
\end{tabularx}

%% file: tables/new_stats.tex
\caption{Comprehensive statistics of \textsc{FollowTable}. $|\mathcal{Q}|$, $|\mathcal{I}|$, and $|\mathcal{T_Q}|$ denote the number of queries, instructions, and candidate tables, respectively. 
\textit{Avg}. Row \& Col represent the average number of rows \& columns per table, 
Hier. means whether containing hierarchical headers.
Relevance density metrics ($|\mathcal{T}^+_\mathcal{Q}|/|\mathcal{Q}|$ and $|\mathcal{T}^+_\mathcal{Q,I}|/|\mathcal{I}|$) represent the average number of relevant tables per query and compliant tables per instruction. 
\textbf{Len (Q) and Len (I) indicate the average number of words in queries and instructions, respectively.} Columns 1.1--2.3 show the distribution of instructions across different categories in our proposed taxonomy.}
\label{tab:new_stats}
\resizebox{\textwidth}{!}{%
	\begin{tabular}{l|ccc|ccc|cc|cc|ccccc}
		\toprule
		\multirow{2}{*}{ Dataset} & \multicolumn{3}{c|}{ Scale} & \multicolumn{3}{c|}{ Table Structure} & \multicolumn{2}{c|}{ Relevance Density} & \multicolumn{2}{c|}{ Input Length} & \multicolumn{5}{c}{ Instruction Scale By Type}                                                                                                                                                         \\
		\cmidrule(lr){2-4} \cmidrule(lr){5-7} \cmidrule(lr){8-9} \cmidrule(lr){10-11} \cmidrule(lr){12-16}
		                          & $|\mathcal{Q}|$     & $|\mathcal{I}|$                & $|\mathcal{T_Q}|$                       & \textit{Avg}. Row                  & \textit{Avg}. Col                              & Hier.           & $|\mathcal{T}^+_\mathcal{Q}|/|\mathcal{Q}|$ & $|\mathcal{T}^+_\mathcal{Q,I}|/|\mathcal{I}|$ & Len (Q) & Len (I)  & C1 & C2 & S1 & S2 & S3  \\
		\midrule
		WQT                       & 300                 & 1386                           & 23,784                                  & 8.13                               & 4.10                                         &   & 39.22                                       & 20.79                                         & 6.14    & 126.53      & 268 & 243 & 297 & 281 & 297  \\
		WTR                       & 60                  & 267                            & 9,546                                   & 5.74                               & 22.47                                        &   & 46.55                                       & 21.22                                         & 2.80    & 126.21      & 50  & 46  & 59  & 54  & 58   \\
		TArX                      & 97                  & 408                            & 11,586                                  & 10.86                              & 5.38                                         &  \ding{51}  & 30.88                                       & 16.89                                         & 5.22    & 43.36       & 69  & 79  & 90  & 83  & 87   \\
		\textsc{IndusTR}          & 216                 & 928                            & 13,258                                  & 8.15                               & 5.33                                         &  \ding{51}  & 33.81                                       & 15.38                                         & 6.95    & 133.55      & 206 & 194 & 211 & 109 & 208  \\
		\bottomrule
	\end{tabular}
}

%% file: sections/6_experiment1.tex
\section{Experiments \& Analysis}
\begin{table*}[th!]
    \input{tables/main_results}
\end{table*}

\subsection{Experimental Setup}

\subsubsection{Evaluated Models}

The goal of the benchmark is to determine how effectively existing retrieval models perform in IFTR. We evaluate models across three categories which listed in Table~\ref{table:full_results}~\footnote{Many Specialized Table Retrievers as noted in \S~\ref{sec:related_tr} do not release either code or model weights, and thus cannot be included in our evaluation.}.

\subsubsection{Implementation Details.} 
For all general retrievers, we serialize tables in HTML or Markdown format. For models with constrained context windows, we truncate tables by preserving the header and a limited number of rows. 

\subsubsection{Metrics}
We adopt a multi-dimensional evaluation approach using the following metrics:

nDCG@10.
We report nDCG@~\cite{nDCG} scores for both \textit{query-only} and \textit{query+instruction} inputs to evaluate general retrieval effectiveness. 
While nDCG effectively reflects topical relevance, it remains agnostic to instruction compliance, failing to distinguish between instruction-compliant and instruction-violating results as long as they are topically relevant.

p-MRR.
To quantify the relative rank change of a designated target document under instructions, we include p-MRR~\cite{weller2025followir}, 
which measures the ratio-based change in reciprocal rank of the gold document between the original and instructed rankings.

IRS.
As our primary metric for instruction adherence, the details of it is in \S~\ref{sec:irs}. 

\subsection{Overview}
Our experiments are designed to answer the following research questions:

\begin{itemize}[leftmargin=1.5em, nosep]
    \item \textbf{RQ1 (Overall Performance):} How do current retrievers perform on the \textsc{FollowTable} benchmark? (\S ~\ref{sec:exp_main})
    \item \textbf{RQ2 (Reranking Gain)}: Can high-capacity cross-encoders compensate for the instruction-following deficiencies of bi-encoder retrievers? (\S~\ref{sec:rerank})
    \item \textbf{RQ3 (Instruction Types)}: Which constraint types pose greater challenges for current models? (\S~\ref{sec:breakdown})
    \item \textbf{RQ4 (Negation Sensitivity):} Are retrievers susceptible to "positive attention bias" when processing negative or exclusionary constraints? (\S~\ref{sec:sensitivity})
    \item \textbf{RQ5 (Metric Robustness):} Does the proposed IRS metric provide a more sensitive and monotonic signal of instruction adherence than traditional IR metrics? (\S~\ref{sec:metric_exp})
\end{itemize}

\subsection{Main Results on \textsc{FollowTable} (RQ1)}
\label{sec:exp_main}

Table~\ref{table:full_results} presents the results of all baselines on the four sub-datasets of \textsc{FollowTable} as well as their average performance.
From the table, we can observe that: 
Table~\ref{table:full_results} presents the results of all baselines on the four sub-datasets of \textsc{FollowTable} as well as their average performance.
From the table, we can observe that: 
\begin{itemize}[leftmargin=1.5em, nosep]
    \item Specialized table retrievers generally underperform general-purpose retrievers across most benchmarks, despite their ability to model table-specific structures (e.g., row–column interactions). 
    Their advantages are largely confined to specific datasets that closely resemble their pre-training data (e.g., WQT), highlighting their limited cross-domain generalization.
    \item While instruction-following retrievers show marginal gains over their non-instructional counterparts in standard ad-hoc TR settings, they significantly outperform all other models in instruction-following capabilities, though substantial room for improvement remains.
    \item Discrepancies observed between high nDCG@10(+$I$) scores and low p-MRR/IRS values suggest that nDCG alone is an insufficient metric for capturing how instructions specifically influence ranking dynamics.
\end{itemize}

\subsection{Effectiveness of Re-rankers (RQ2)}
\label{sec:rerank}
\begin{table}[t]
	\input{tables/reranker}
\end{table}
To address RQ2, we evaluated several SOTA re-rankers, as detailed in Table \ref{table:rerank_results}.
Since re-rankers are used to refine initial retrieval results, we re-rank the top-100 candidates produced by Qwen3-Emb-0.6B, which offers a strong efficiency–performance trade-off. 
We also report the runtime of each re-ranker relative to Qwen3-Emb-0.6B as a reference, with all experiments conducted on a single NVIDIA RTX 4090 GPU.

Our empirical analysis yields several key insights:
\begin{itemize}
    \item Specialized vs. General Re-rankers: Table-specialized cross-encoders are consistently outperformed by general-purpose re-rankers. This suggests that the latter benefit significantly from the superior semantic understanding and reasoning capabilities inherent in LLMs.
    However, even SOTA LLM-based re-rankers struggle in the list-wise setting to reorder tables based on whether they fully comply with the given instructions.
    \item Performance vs. Latency Trade-offs: While general re-rankers substantially outperform bi-encoder retrievers in terms of IRS, they incur a computational cost that is 1–2 orders of magnitude higher.
\end{itemize}

\subsection{Performance Analysis Across Instruction Types (RQ3)}
\label{sec:breakdown}
\begin{figure}
    \centering
    \includegraphics[width=0.98\linewidth]{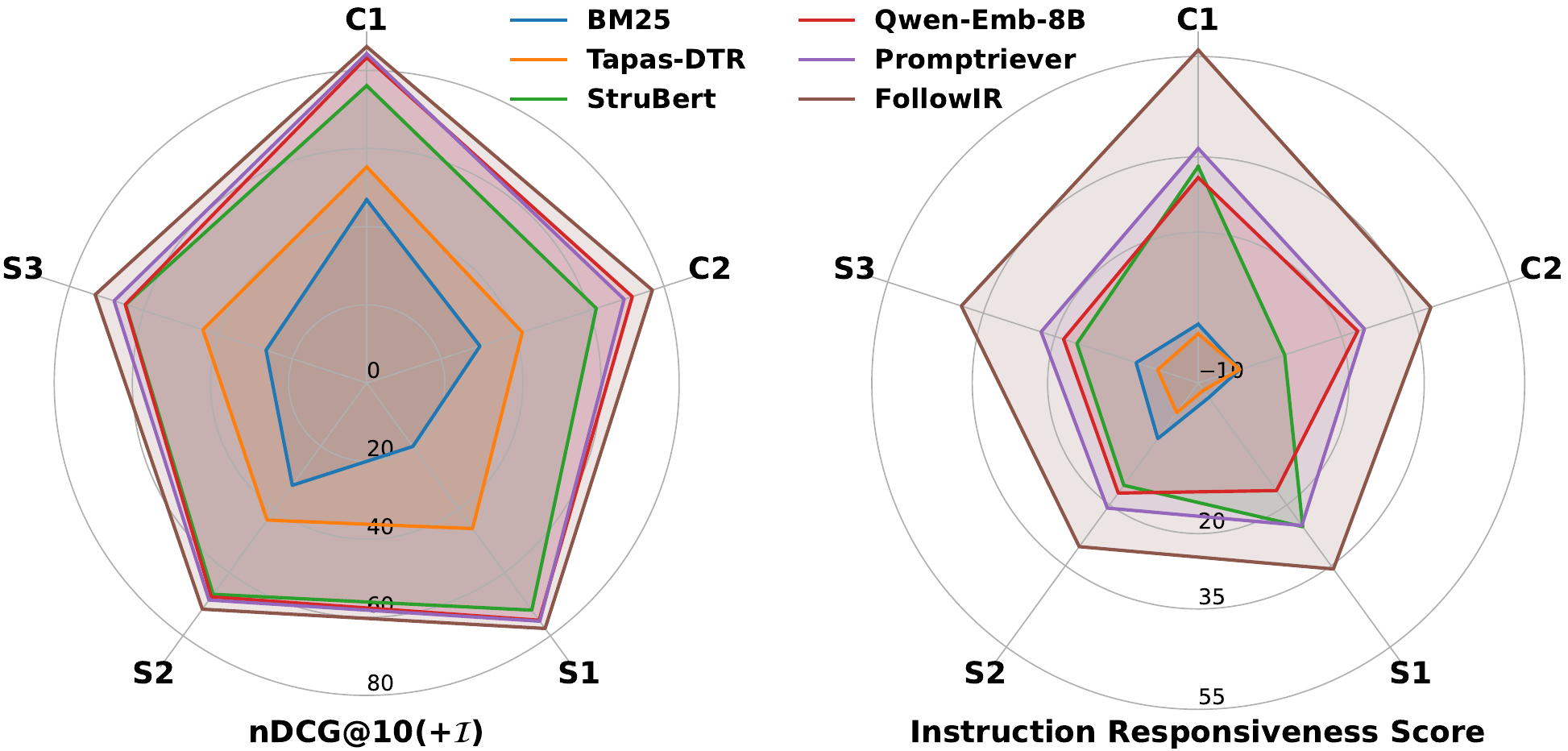}
    \vspace{-0.2cm}
    \caption{Performance comparison about nDCG@10 on instructions (left) and IRS (right) across different instruction constraint types.}
    \vspace{-0.3cm}
    \label{fig:radar}
\end{figure}
RQ3 investigates which types of instructions pose the greatest challenges for existing retrievers. 
We evaluate a diverse set of retrievers and re-rankers across five instruction categories, and summarize the results in Fig.~\ref{fig:radar}. 
A clear observation is that strong general-purpose re-rankers substantially outperform other models across all five instruction types.

From the perspective of instruction difficulty, instruction-aware retrievers and re-rankers with stronger capabilities exhibit more systematic and stable behavior. Among the five categories, C1 (Semantic Boundary Constraint) is generally the easiest to satisfy, while C2 (Exclusive Topic Constraint), S1 (Attribute-centric Structural Constraint), and S3 (Granularity-centric Structural Constraint) pose moderate challenges. Among all categories, the Entity-centric Structural Constraint (S2) is the most difficult, suggesting that aligning retrieval behavior with entity-level structural requirements remains a major bottleneck for current retrievers.
Notably, while all models peak at C1, StruBERT uniquely shows a relative strength in S1 compared to its performance on other structural constraints. This indicates that its table-specific pre-training confers superior column-level semantic understanding over general-purpose re-rankers.

\subsection{Sensitivity to Negated Entities (RQ4)}
\label{sec:sensitivity}
Some IFIR studies~\cite{weller2025followir, InfoSearch} have highlighted a critical vulnerability in dense retrievers: the sensitivity to negated keywords. 
While modern retrievers are increasingly adept at capturing positive semantic alignment, their ability to process negative constraints (e.g., "excluding," "without," "never") remains questionable. 
We hypothesize that many models suffer from a "Positive Attention Bias," where the presence of a negated entity in the instruction inadvertently increases the relevance score of documents containing that entity.

We sample 50 query–instruction pairs and use the corresponding set of instruction-violating tables, denoted as $\mathcal{T}^{-}_{qi}$.
Within this set, we further construct a subset of trap tables $\mathcal{T}_{\text{trap}} \subseteq \mathcal{T}^{-}_{qi}$, which violate the instruction specifically due to the presence of explicitly negated entities.
These trap tables are used to evaluate model sensitivity to negation and exclusionary constraints.

To quantify the extent to which a model is misled by negated entities, we define the Negation Failure Rate (NFR) to quantify the extent to which a model is misled by negated entities in instructions, by measuring how often trap tables $t\in\mathcal{T}_{trap}$ are promoted after instruction conditioning.
Formally, given a query $q$ and an instruction $i$, $\pi_q$ and $\pi_{q,i}$ denote the ranked lists
produced without and with instruction conditioning, respectively.
The NFR is defined as:
\[
\mathrm{NFR}
=
\frac{1}{|\mathcal{T}_{trap}|}
\sum_{t \in \mathcal{T}_{trap}}
\mathbb{I}
\bigl(
\mathrm{r}(t,\pi_{q,i})
<
\mathrm{r}(t,\pi_q)
\bigr),
\]
where $\mathrm{r}(t,\pi)$ denotes the rank position of table $t$ in the ranked list $\pi$,
with smaller values indicating higher ranks.

\begin{table}[t]
    \input{tables/nfr}
\end{table}

As shown in Table~\ref{table:nfr},three representative models exhibits a high NFR, suggesting that the "Frequency Bias" of negated tokens often outweighs the logical semantics of the instruction.

\begin{figure}[t]
  \centering
  \includegraphics[width=0.8\linewidth]{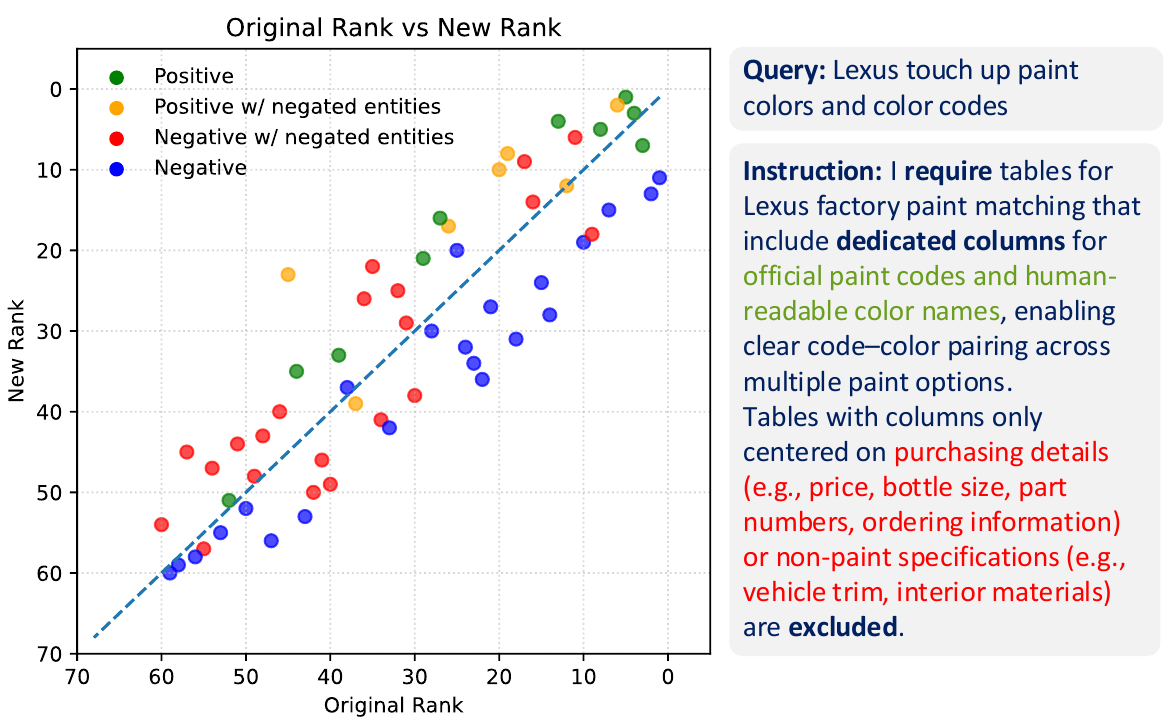}
  \vspace{-0.3cm}
\caption{Rank shift analysis of Promptriever on a sample query with a type S1 instruction. The right panel shows the query and instruction, where red words indicate excluded tables and the green indicate desired tables.}
  \label{fig:scatter}
  \vspace{-0.2cm}
\end{figure}

\textbf{Case Study: Rank Shift Analysis.} 
To further illustrate this failure mode, we conduct a rank shift analysis case study and present the corresponding scatter plot in Fig.~\ref{fig:scatter} for SOTA Promptriever on a representative query selected from the WQT, paired with a Type S1 (Attribute-centric Structural Constraint) instruction.
For this query, we construct a candidate set of 60 tables that are topically aligned with the query and then categorize them into four groups. 
The first two groups (\textcolor[rgb]{0,0.5,0}{green} and \textcolor[rgb]{1,0.647,0}{orange} points) satisfy the instruction constraints; however, the second group still contains a subset of entities that the user explicitly indicated as unnecessary.
The remaining two groups violate the instruction: the third group (\textcolor[rgb]{1,0,0}{red} points) explicitly includes negated entities, while the fourth group (\textcolor[rgb]{0,0,1}{blue} points) fails to meet the instruction requirements in other ways.
As shown in Fig.~\ref{fig:scatter}, even the SOTA retrieval model exhibits limited instruction-following capability, as reflected by the generally small magnitude of rank changes across all documents.
More importantly, the model is highly sensitive to the presence of negated entities: a substantial portion of the \textcolor[rgb]{1,0.647,0}{orange} and \textcolor[rgb]{1,0,0}{red} points experience rank promotion. 
This indicates that explicitly negated entities can still exert a strong positive influence on retrieval scores, revealing the model’s insufficient handling of negation constraints.

\subsection{Metric Validity (RQ5)}
\label{sec:metric_exp}
\begin{figure}
    \centering
    \includegraphics[width=0.97\linewidth]{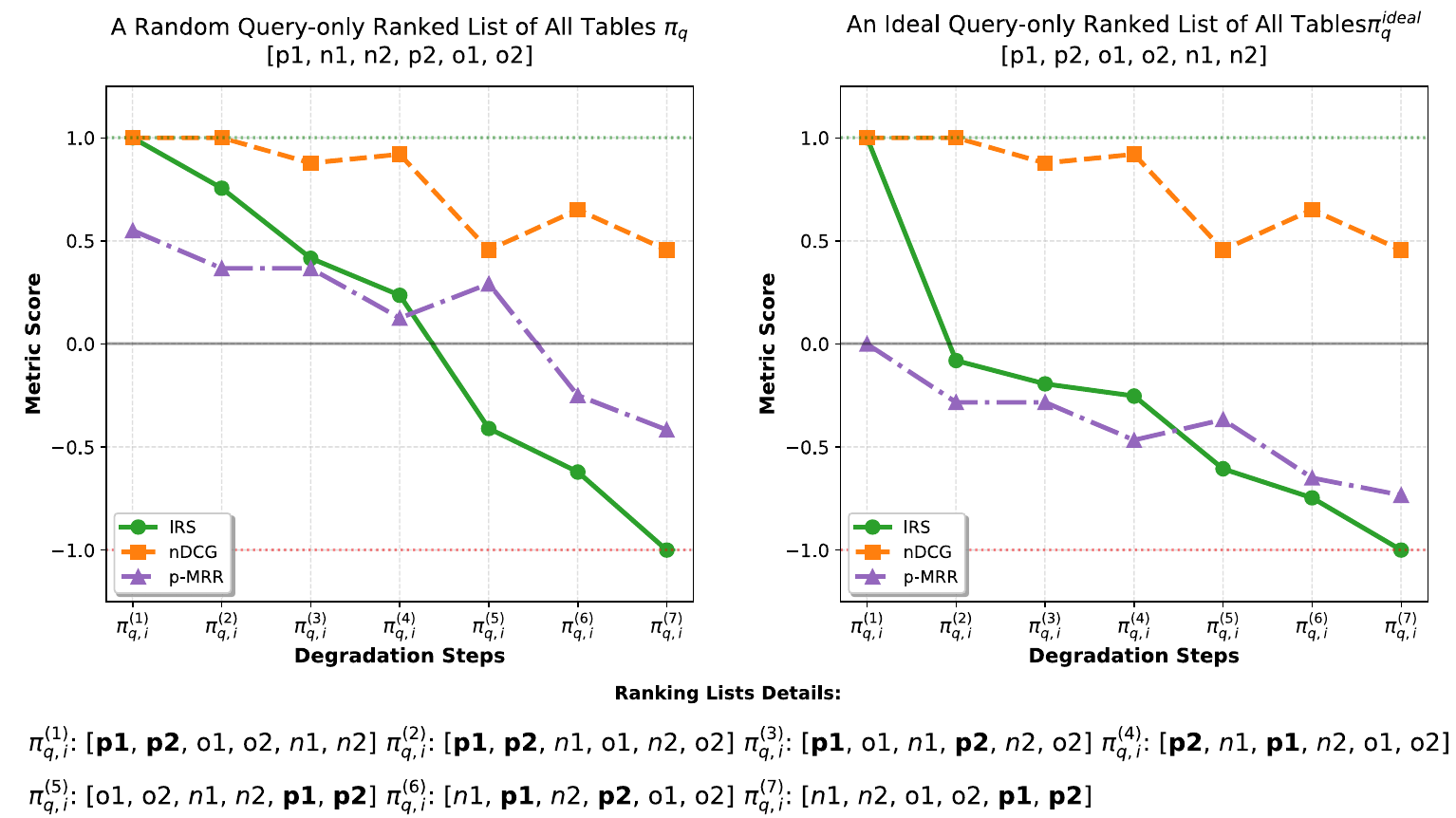}
    \vspace{-0.2cm}
\caption{Comparison of nDCG, p-MRR, and IRS under controlled ranking degradation. IRS exhibits strict monotonicity as instruction compliance decreases ($\pi_0 \to \pi_6$). In contrast, nDCG and p-MRR fail to provide a reliable signal.}
    \vspace{-0.3cm}
    \label{fig:metric_validity}
\end{figure}
To evaluate whether metrics faithfully reflect instruction-following, we construct a controlled scenario (Fig.~\ref{fig:metric_validity}) using a toy corpus $\{p_1, p_2, n_1, n_2, o_1, o_2\}$. 
Here,  $n$ documents are topically relevant to query $q$ but violate instruction $i$, while $p$ documents satisfy both. 
We simulate two query-only baselines (random vs. ideal) and progressively degrade the rankings from $\pi_0$ to $\pi_6$ by promoting instruction-violating ($n$) or irrelevant ($o$) documents.

As shown in Fig.~\ref{fig:metric_validity}, IRS exhibits strict monotonicity, accurately reflecting the diminishing degree of instruction compliance.
In contrast, nDCG lacks sensitivity to instruction-induced conditional relevance, often assigning identical scores to different adherence levels (e.g., $\pi_1$ vs. $\pi_2$). Furthermore, nDCG yields a misleadingly high score near 0.5 even for the poorest ranking results, whereas both p-MRR and IRS correctly produce values below zero in such cases. 
p-MRR similarly displays limited discriminative power (e.g., $\pi_2$ vs. $\pi_3$ or $\pi_4$ vs. $\pi_5$), as its score is dominated solely by the ranks of $n_i$ documents, leading to irregular fluctuations that fail to provide a stable signal of ranking quality. 
The most critical failure of p-MRR occurs when the query-only ranking is already optimal (right panel). 
In this case, maintaining the same optimal ranking under $(q,i)$ yields a zero score for p-MRR, as it perceives no observable ranking change. 
Conversely, both nDCG and IRS correctly assign perfect scores, accurately acknowledging full instruction satisfaction. This demonstrates that only IRS provides a discriminative and faithful assessment across the entire instruction-following process.

\label{sec:case_study}

%% file: tables/main_results.tex
\caption{Comprehensive evaluation across four datasets and the overall average. We report nDCG@10 for original queries (\textbf{Q}) and queries with instructions (\textbf{+I}), plus \textbf{p-MRR} and \textbf{IRS} (both scaled to [-100, +100]). Best results in bold.}
\vspace{-0.1cm}
\label{table:full_results}
\centering
\setlength{\tabcolsep}{2.3pt}
\renewcommand{\arraystretch}{1.15}
\resizebox{\textwidth}{!}{%
	\begin{tabular}{@{}l|c@{\hspace{11pt}}ccc|c@{\hspace{11pt}}ccc|c@{\hspace{11pt}}ccc|c@{\hspace{11pt}}ccc|c@{\hspace{11pt}}ccc@{}}
		\toprule
		\multirow{3}{*}{\textbf{Model}}
		                                         & \multicolumn{4}{c|}{\textbf{WQT}} & \multicolumn{4}{c|}{\textbf{WTR}} & \multicolumn{4}{c|}{\textbf{TArX}} & \multicolumn{4}{c|}{\textbf{IndusTR}} & \multicolumn{4}{c}{\textbf{Average}}                                                                                                                                                                                                    \\

		\cmidrule(lr){2-5} \cmidrule(lr){6-9} \cmidrule(lr){10-13} \cmidrule(lr){14-17} \cmidrule(l){18-21}

		                                         & \multicolumn{2}{c}{nDCG }         & \multirow{2}{*}{p-MRR }           & \multirow{2}{*}{IRS}
		                                         & \multicolumn{2}{c}{nDCG }         & \multirow{2}{*}{p-MRR }           & \multirow{2}{*}{IRS }
		                                         & \multicolumn{2}{c}{nDCG }         & \multirow{2}{*}{p-MRR }           & \multirow{2}{*}{IRS }
		                                         & \multicolumn{2}{c}{nDCG }         & \multirow{2}{*}{p-MRR }           & \multirow{2}{*}{IRS }
		                                         & \multicolumn{2}{c}{nDCG }         & \multirow{2}{*}{p-MRR }           & \multirow{2}{*}{IRS }                                                                                                                                                                                                                                                                                                \\

		\cmidrule(lr){2-3} \cmidrule(lr){6-7} \cmidrule(lr){10-11} \cmidrule(lr){14-15} \cmidrule(lr){18-19}

		                                         & Q                                 & +I                                &                                    &                                       & Q                                    & +I   &      &               & Q    & +I   &      &               & Q             & +I            &               &               & Q             & +I            &               &               \\
		\midrule

		\multicolumn{21}{c}{\textit{\textbf{No-Instruction Retrievers}}}                                                                                                                                                                                                                                                                                                                                                                        \\
		\midrule
		BM25~\citep{bm25}                        & 65.3                              & 30.3                              & 2.4                                & -1.7                                  & 52.5                                 & 12.2 & 3.8  & 2.5           & 60.7 & 44.4 & -8.0 & 5.7           & 62.7          & 36.9          & 0.8           & 8.3           & 60.3          & 31.0          & -0.5          & 3.7           \\
		BGE-Large-v1.5~\citep{bge_large}         & 73.6                              & 46.9                              & 2.8                                & 0.1                                   & 84.7                                 & 44.5 & 6.4  & -0.5          & 77.7 & 59.8 & -8.2 & 9.0           & 76.7          & 50.9          & -5.9          & 9.2           & 78.2          & 50.5          & -1.2          & 4.5           \\
		E5-Large-v2~\citep{e5_large}             & 70.6                              & 41.1                              & 3.2                                & -1.9                                  & 79.9                                 & 43.5 & 7.3  & 2.5           & 77.4 & 49.9 & -0.1 & 3.8           & 72.2          & 45.6          & -3.6          & 9.3           & 75.0          & 45.0          & 1.7           & 3.4           \\
		ReasonIR (8B)~\cite{shao2025reasonir}    & 69.9                              & 40.2                              & -2.4                               & -2.0                                  & 76.9                                 & 42.1 & -3.6 & -1.7          & 75.1 & 54.5 & -9.9 & 7.3           & 66.8          & 44.8          & -11.3         & 8.7           & 72.2          & 45.4          & -6.8          & 3.1           \\
		\midrule

		\multicolumn{21}{c}{\textit{\textbf{No-Instruction Specialized Table Retrievers}}}                                                                                                                                                                                                                                                                                                                                                      \\
		\midrule
		Tapas-DTR~\cite{nq_tables}               & 75.2                              & 50.7                              & 5.8                                & -1.1                                  & 78.8                                 & 50.8 & 0.6  & -6.7          & 70.3 & 45.5 & -3.3 & -5.1          & 60.2          & 40.4          & 2.2           & 1.6           & 71.1          & 46.9          & 1.3           & -2.8          \\
		Table-GTR~\cite{gtr}                     &	67.1	&	45.2	&	2.6	&	0.3                                  &	54.7	&	41.1	&	-1.2	&	0.5           &	65.1	&	39.1	&	-4.2	&	-2.1           &	57.2	&	38.1	&	-7.0	&	-1.5           &	61.0	&	40.9	&	-2.5	&	-0.7\\
		Table-DPR~\cite{TR_Necessitate}          & 77.0                                & 54.2                              & -2.3                               & -1.6                                  & 53.3                                 & 20.0   & -8.7 & -9.1          & 73.2 & 51.4 & -2.9 & -2.5          & 70.6          & 45.6          & -1.9          & -5.6          & 68.5        & 42.8          & -4.0         & -4.7          \\
		Birdie~\cite{Birdie}                     &	\textbf{79.3}	&	63.1	&	3.6	&	2.1 &	74.2	&	57.3	&	6.4	&	1.5 &	67.2	&	55.6	&	-1.2	&	2.8 &	64.1	&	49.6	&	-5.6	&	2.3 &	71.2	&	56.4	&	0.8	&	2.2 \\
		\midrule

		\multicolumn{21}{c}{\textit{\textbf{Instruction-Following Retrievers}}}                                                                                                                                                                                                                                                                                                                                                                 \\
		\midrule
		Instructor-XL (1.5B)~\cite{one_embedder} & 73.0                              & 47.4                              & 1.7                                & 1.3                                   & 83.4                                 & 52.9 & 3.6  & 1.6           & 78.8 & 56.2 & -4.5 & 8.4           & 72.0          & 47.9          & -4.2          & 9.9           & 76.8          & 51.1          & -0.9          & 5.3           \\
		E5-Mistral (7B)~\citep{e5_mistral}       & 76.2                              & 58.6                              & 2.0                                & 6.2                                   & 87.2                                 & 69.3 & 1.9  & 9.0           & 87.4 & 68.0 & 2.2  & 14.1          & 74.3          & 57.1          & -0.6          & 17.1          & 81.3          & 63.3          & 1.4           & 11.6          \\
		GritLM (7B)~\citep{gritlm}               & 76.3                              & 58.0                              & 2.9                                & 6.7                                   & 86.0                                 & 66.7 & 5.8  & 10.7          & 85.5 & 69.5 & 1.8  & 14.7          & \textbf{77.3} & 60.3          & 2.1           & 19.3          & 81.3          & 63.6          & 3.1           & 12.9          \\

		OpenAI v3 Large                          & 75.7                              & 54.6                              & 2.4                                & 4.3                                   & 84.1                                 & 60.2 & 3.0  & 7.8           & 87.0 & 66.2 & 0.4  & 11.7          & 74.4          & 52.9          & -0.3          & 13.7          & 80.3          & 58.5          & 1.4           & 9.4           \\
		Qwen3-Emb (0.6B)~\citep{zhang2025qwen3}  & 75.4                              & 59.6                              & 4.2                                & 9.6                                   & 80.4                                 & 65.7 & 5.4  & 15.5          & 84.2 & 65.9 & 4.6  & 16.4          & 72.0          & 60.1          & 0.3           & 21.1          & 78.0          & 62.8          & 3.6           & 15.6          \\
		Qwen3-Emb (8B)~\cite{zhang2025qwen3}     & 77.1                              & 66.8                              & 7.5                                & 14.5                                  & \textbf{88.3}                                 & 75.4 & 7.1  & 17.6          & \textbf{88.8} & \textbf{77.6} & 14.2 & 24.7          & 76.2          & 70.7          & \textbf{11.5} & 31.8          & \textbf{82.6} & 72.6          & 10.1          & 22.2          \\
		Promptriever (7B)~\cite{Promptriever}    & 75.1                              & \textbf{69.1}                              & \textbf{8.2}                                & \textbf{18.3}                         & 82.3                                 & \textbf{76.4} & \textbf{10.6} & \textbf{23.7} & 83.7 & 76.5 & \textbf{15.1} & \textbf{28.7} & 69.9          & \textbf{71.2} & 8.0           & \textbf{34.2} & 77.7          & \textbf{73.3} & \textbf{10.5} & \textbf{26.2} \\

		\bottomrule
	\end{tabular}%
}

%% file: tables/reranker.tex
\caption{Performance evaluation of re-ranking models on top-100 candidates retrieved by Qwen3-Emb-0.6B.}
\label{table:rerank_results}
\centering
\setlength{\tabcolsep}{3.5pt}
\renewcommand{\arraystretch}{1.25}

\resizebox{.95\columnwidth}{!}{%
\begin{tabular}{@{}lcccccc@{}}
\toprule
\multirow{2}{*}{\textbf{Model}} 
& \multirow{2}{*}{\textbf{Type}} 
& \multicolumn{4}{c}{\textbf{Average Performance}} 
& \multirow{2}{*}{\textbf{Time}} \\
\cmidrule(l){3-6}
& & nDCG(Q) & nDCG(+I) & p-MRR & IRS & \\
\midrule

Qwen3-Emb-0.6B~\cite{zhang2025qwen3} 
& Retriever 
& 78.0 & 62.8 & 3.6 & 15.6
& 1.0$\times$  \\
\midrule
\midrule

\multicolumn{7}{c}{\textit{\textbf{Table-specialized Cross-Encoders}}} \\
\midrule
StruBERT~\cite{strbert} 
& Point-wise 
& 67.5 & 68.4 & -2.34 & 19.7 
& 10$\times$ \\
\midrule

\multicolumn{7}{c}{\textit{\textbf{General Rerankers}}} \\
\midrule
FollowIR (7B)~\cite{weller2025followir} 
& Point-wise
& 80.6 & 77.5 & 18.0 & 43.0
& 122$\times$ \\
Rank1 (7B)~\cite{rank1} 
& Point-wise  
& 83.1 & 79.0 & 18.3 & 46.3
& 403$\times$ \\
RankZephyr (7B)~\cite{pradeep2023rankzephyr} 
& List-wise 
& 84.6 & 65.8 & 5.9 & 23.1 
& 83$\times$ \\
\midrule
\multicolumn{7}{c}{\textit{\textbf{SOTA LLMs}}} \\
\midrule
Gemini-3-Pro~\cite{deepmind_gemini3pro_modelcard} 
& List-wise
& 92.9 & 92.4 & 21.6 & 58.1 
& / \\
\bottomrule
\end{tabular}%
}

%% file: tables/nfr.tex
\caption{Negation Failure Rate (NFR) of different retrieval and reranking models.
NFR measures the proportion of instruction-violating trap tables that are promoted in the ranking after instruction conditioning.
Lower values indicate better robustness to negated entities.}
\vspace{-0.3cm}
\centering
\resizebox{0.95\columnwidth}{!}{%

\begin{tabular}{ccc}
    \toprule
    Model & Type & NFR \\
    \midrule
    BM25 & No-Instruction Retrievers & 0.82 \\
    E5-Large-v2 & No-Instruction Retrievers & 0.71 \\
    Birdie & No-Instruction Specialized Table Retrievers & 0.73 \\
    Promptriever & Instruction-Following Retrievers & 0.61 \\
    FollowIR & General Re-rankers & 0.35 \\
    \bottomrule
\end{tabular}
}

\vspace{-0.4cm}
\label{table:nfr}

%% file: sections/6_experiment2.tex
\section{Discussion}
\label{sec:discussion}

\subsection{General vs. Specialized Table Retrievers}
In ad-hoc TR, specialized table retrievers only excel in-domain by modeling table-specific structures but generalize poorly to unseen data. 
Conversely, general-purpose retrievers—particularly those leveraging LLM backbones on linearized tables—demonstrate superior zero-shot robustness. 
However, this robustness does not extend to IFTR; general models excel at topical relevance but struggle with table-centered content and schema-grounded constraints. 
This suggests the limitation is not structural encoding capability, but a lack of instruction-driven signals. 
Consequently, rather than developing narrow specialized architectures, a more effective strategy is to augment robust general-purpose encoders with instruction-tuning to align natural language constraints with tabular structure.

\subsection{Retrieval vs. Re-ranking in IFTR}
Consistent with prior work, cross-encoder re-rankers outperform retrievers in validating complex IFTR constraints at prohibitive latency costs. However, the unexpected efficacy of instruction-aware retrievers (e.g., Promptriever) suggests that instruction adherence is not exclusive to the re-ranking stage. We argue that incorporating IFTR-specific data into dense retriever training offers a viable path to reduce reliance on expensive re-ranking. This approach shifts the computational burden upstream, achieving a more favorable balance between constraint satisfaction and system efficiency.

%% file: sections/8_conclusion.tex
\section{Conclusion}

We introduced \textsc{FollowTable}, the first benchmark for IFTR. 
Our analysis reveals that while current retrievers handle topical relevance well, they struggle with the schema and content constraints within instructions. 
Furthermore, our proposed IRS metric provides a more robust measure of instruction adherence than traditional IR metrics. 
These findings highlight the critical need to train future general or specialized table retrievers to align user instructions with both tabular content and schema semantics.